\documentclass[twocolumn,10pt]{article}
\usepackage[a4paper,margin=1.9cm,columnsep=0.7cm]{geometry}
\usepackage{etoolbox}
\usepackage{xspace}

\usepackage[numbers]{natbib}

\usepackage{amsmath,amssymb,amsfonts}

\usepackage{graphicx}
\usepackage{textcomp}
\usepackage{pdflscape}
\usepackage[ruled,vlined,linesnumbered]{algorithm2e}

\usepackage{algpseudocode}
\usepackage[loadonly]{enumitem}
\usepackage{makecell}
\usepackage{arydshln}
\usepackage{booktabs, array, multirow, longtable}
\usepackage{adjustbox}
\usepackage{multirow, multicol}
\usepackage{subcaption}
\usepackage{stfloats}
\usepackage{comment}
\usepackage{hyperref}
\usepackage{url}
\usepackage{threeparttable} 
\usepackage{threeparttablex} 
\usepackage{longtable}
\usepackage{tabularx}   
\usepackage{booktabs}   

\usepackage{setspace}

\usepackage{array}      
\usepackage{makecell}   
\usepackage{xparse}   
\usepackage{caption}  
\usepackage{ltablex} 
\usepackage{float}
\usepackage{xcolor}
\newcommand{\credit}[1]{}   
\newcommand{\printcredits}{}

\begin{document}

\hypersetup{
  pdftitle={FedASAP: Activation Statistics-driven Structured Adaptive Pruning for Efficient Personalized Federated Learning for Lesion Segmentation on brain MRI},
  pdfauthor={Karan R. Bagri, Tarun K. Garg, Vaanathi Sundaresan}
}

\twocolumn[{%
  \begin{center}
    {\LARGE\bfseries FedASAP: Activation Statistics-driven Structured Adaptive Pruning for Efficient Personalized Federated Learning for Lesion Segmentation on brain MRI\par}
    \vspace{1.2em}
    {\large
      Karan R. Bagri\textsuperscript{1,\,$\ast$}\quad
      Tarun K. Garg\textsuperscript{2,\,$\ast$}\quad
      Vaanathi Sundaresan\textsuperscript{1,\,$\dagger$}\par}
    \vspace{0.8em}
    {\normalsize
      \textsuperscript{1}Department of Computational and Data Sciences, Indian Institute of Science, Bengaluru 560012, Karnataka, India\par
      \textsuperscript{2}Department of IISc Mathematics Initiative, Indian Institute of Science, Bengaluru 560012, Karnataka, India\par}
    \vspace{0.5em}
    {\footnotesize
      \textsuperscript{$\ast$}Karan R. Bagri and Tarun K. Garg contributed equally to this work.\quad
      \textsuperscript{$\dagger$}Corresponding author: \href{mailto:vaanathi@iisc.ac.in}{vaanathi@iisc.ac.in}\par}
  \end{center}
  \vspace{1em}
  \begin{center}
  \begin{minipage}{0.92\textwidth}
    \noindent\textbf{Abstract.}\;
In medical imaging, developing robust deep learning models requires data from various domains. However, regulatory policies protecting patient privacy restrict data sharing. Federated learning (FL) addresses this by enabling collaborative model training without centralizing data. Yet, data heterogeneity across client centers requires personalized models to enhance performance. Clients with limited resources may struggle to efficiently train or deploy large deep neural networks. Adaptive model pruning tackles both challenges by reducing model size while enabling personalized FL adaptation. Although research has investigated adaptive pruning methods in FL, their effectiveness on dense prediction tasks like segmentation remains unexplored. To address this gap, we propose Activation Statistics-driven structured Adaptive Pruning (FedASAP), which uses activation-based features to guide filter removal for each client. By learning a lightweight classifier on per-filter activation statistics, FedASAP refines importance-score rankings and produces compact, personalized segmentation models in heterogeneous federated settings. Our results show the approach's effectiveness by achieving better Dice scores of 0.796 and 0.743, than state-of-the-art, while reducing parameter count by 45\% and 73\% on two multi-centric brain pathology segmentation datasets: Federated Tumour Segmentation (FeTS) and White Matter Hyperintensities (WMH), respectively.\par
    \vspace{0.6em}
  \end{minipage}
  \end{center}
  \vspace{1.5em}
}]

\section{Introduction}
\label{sec:introduction}
Deep learning (DL) models for medical imaging tasks are limited by small datasets and resource constraints. Data sharing across institutions is restricted by regulatory policies (e.g. GDPR~\cite{ref_gdpr} and HIPPA~\cite{ref_hipaa}). Federated learning (FL) framework enables training models on data from multiple sources (clients) without data centralization \cite{ref_fedavg}. This process involves local training where models are trained on each client's data, followed by aggregation at a central server to produce an improved global model. The global model is then distributed to clients. These steps form a training round and repeat until convergence. A key challenge in FL is statistical data heterogeneity between clients~\cite{ref_survey,ref_qu2021}, often leading to client model divergence. In these cases, the locally‑trained models drift away from each other and the intended global optimum. Due to local data distribution variations, unchecked divergence can impair learned representations, resulting in federated models often showing slower convergence and degraded performance. Researchers have explored regularization techniques ~\cite{ref_fedprox,ref_fedmax,ref_moon} that mitigate model drift and stabilize training, and data augmentation ~\cite{ref_dataaug,ref_fedcg,ref_gansynthesis,ref_stablediffusion,ref_fedda} methods to enhance dataset diversity. Personalization strategies~\cite{ref_fedbn,ref_fedrep,ref_fedlc,ref_feddp,ref_fedinout} have been developed to adapt models to client‐specific characteristics, improving performance in non-independent and identically distributed (non‐IID) settings. These techniques enhance FL robustness but often add complexity or computational overhead. 

Healthcare facilities, especially in resource‐constrained settings, struggle with limited infrastructure and high energy costs. Smaller models reduce energy consumption and carbon footprint while enabling deployment on resource-constrained environments. Combining robust FL strategies with model compression techniques like pruning can lead to efficient, high-performing models that are environmentally friendly across diverse healthcare settings. Pruning removes less important weights from a neural network, reducing size and speeding up inference while often maintaining accuracy. By iteratively pruning parameters with low importance (determined using local activation magnitudes~\cite{ref_mag1,ref_mag2,ref_mag3,ref_mag4} or gradient based saliency~\cite{ref_snip,ref_synflow,ref_grasp}), clients can refine architectures to retain weights critical for their specific characteristics. 

Pruning can involve removal of individual weights (unstructured pruning) or entire structures like neurons, channels, filters or layers (structrued pruning). Recent studies have explored pruning in federated settings to reduce communication overhead. These approaches often adopt unstructured sparsity~\cite{ref_autoflip, ref_prunefl, ref_feddst, ref_fedmef}. However, most methods are developed for classification tasks and do not translate well to dense prediction problems like segmentation. Often, pruning of weights/filters is based on gradient-based importance scores or mean activation values, which does not provide a comprehensive task-relevant criteria for efficient pruning and robust performance, especially for segmentation. Moreover, unstructured pruning results in sparse masks that are inefficient on standard hardware, limiting deployment in resource constrained settings.

To overcome challenges posed by existing pruning approaches, we propose FedASAP: a two-stage, activation statistics-driven framework for structured model compression in federated segmentation. In the first stage, each client ranks filters by importance and identifies unimportant filters as pruning candidates. Subsequently, a lightweight Generalized Linear Model (GLM), trained on activation statistics, performs the final selection of filters to prune or retain. By removing entire filters and dependent parameters and communicating minimal information regarding newly pruned filters, our method achieves significant model compression and improved segmentation accuracy under heterogeneous federated settings (including low-data clients), while adding minimal communication overhead to training. Our contributions are summarized as follows:

\begin{itemize}
    \item \textbf{Two step adaptive pruning for FL segmentation:} We introduce a two-step structured pruning framework within FL for segmentation tasks. We initially use a gradient-based importance score. Then, we use a lightweight classifier to select filters to be pruned based on per-filter task-relevant activation features. This provides a flexible and trainable decision-making process in an explainable manner.
    \item \textbf{Reduced info upload for minimal communication overhead:} Unlike existing methods, to minimize communication, clients send only the set of indices of filters that are pruned in each round, drastically reducing upload payload compared to transmitting full masks. 
    \item \textbf{Computational efficiency:} We demonstrate that our  pruning approach yields consistent real-world gains, achieving faster inference times, reduction in peak memory footprint, decrease in overall FLOPs.
    \item \textbf{Structured pruning to minimize divergence: } Our structured pruning strategy acts as a form of implicit regularization, reducing client-level divergence by removing unimportant filters, as evident from the comparisons with earlier FL methods.
    \item \textbf{Comparison with a lesion size-aware thresholding strategy:} In addition to the state-of-the-art comparison, we also design a layer-wise adaptive thresholding (LAT) strategy as an ablation baseline to validate the effectiveness of FedASAP on datasets with different lesion sizes.
\end{itemize}

\section{Related Works}

Earlier FL techniques such as  FedAvg~\cite{ref_fedavg} (weighted averaging of weights suring aggregation) set the foundation for federated learning. Later, methods like FedProx~\cite{ref_fedprox}, MOON~\cite{ref_moon} and FedDyn~\cite{durmus2021federated} have shown superior performance compared to FedAvg~\cite{ref_fedavg}. FedProx adds proximal regularization to penalize deviations between local and global models, while MOON uses contrastive loss to align representations and minimize drift and FedDyn uses a per-client dynamic regularizer to adapt local objectives so local minima align with the global one. Such methods produce only a single global model, which may underfit clients with highly skewed data distributions. Personalization addresses this limitation by bridging the gap between limited local data and the diverse patterns that a global model alone cannot capture, leading to faster convergence and improved performance. In earlier personalization strategies, fine-tuning enables clients to adjust the global model to their local data after federated training. Methods involve sharing certain parameters with the global model while keeping others private. For instance, FedBN~\cite{ref_fedbn} addresses feature shifts through local batch normalization updates. FedDP~\cite{ref_feddp} preserves query embedding privacy with inconsistency-guided calibration. FedRep~\cite{ref_fedrep} separates shared features from private prediction heads. LC-Fed~\cite{ref_fedlc} uses contrastive client embeddings for personalized selection. GRACE~\cite{ref_grace} employs client-side feature alignment and server-side consistency re-weighting to enhance personalization. While the above methods focus solely on client-side personalization, pruning can also be used to create lightweight models that align with individual client data distributions. This section explores previous work on pruning methods in neural networks and pruning techniques within federated learning. 

\subsection{Neural network pruning}

Early neural network pruning focused on unstructured sparsity to eliminate individual low-magnitude weights. The Deep Compression pipeline~\cite{ref_mag1} pruned weights with small absolute values, using weight sharing and Huffman coding for compact models. SNIP~\cite{ref_snip} computes saliency scores using loss gradients at initialization to prune low-saliency connections in one shot before training. Methods like lottery ticket hypothesis~\cite{ref_mag3}, SynFlow~\cite{ref_synflow}, and GRASP~\cite{ref_grasp} perform unstructured pruning by removing individual weights, rather than structured components like entire filters or layers. While effective in reducing the number of non-zero parameters, this pruning requires specialized hardware or libraries for actual speed-ups, limiting practical benefits.

More recently, structured pruning addresses these deployment limitations by explicitly reducing network size through removal of entire structural components. STAMP~\cite{ref_stamp}, for instance, reduces U-Net models using filter pruning and channel-wise targeted dropout during training, even improving performance, particularly in low-data regimes. In general, unstructured pruning solely considers individual weight importance, while structured pruning enforces grouping among weights. Hence, achieving consistency in identifying unimportant weight groups is more challenging and requires better training, as important parameters may be removed within an unimportant group~\cite{ref_single_shot,ref_llmsurgeon,ref_surveyprune}. 

Several recent approaches rank filters using activation-based criteria rather than weight magnitude. However, so far, only a couple of features (e.g., feature-map texture,  entropy~\cite{liu2023filter},~\cite{yao2021deep}), and often times only a single feature (e.g., mean~\cite{zhao2022iterative}, entropy~\cite{luo2017entropy}) have been used to replace weight-based importance. Replacing the gradient-based importance score (usually used in pruning) with a single or limited set of handcrafted activation statistics may not be comprehensive and does not capture the task-relevant activation characteristics sufficiently. Integrating both gradient-based scores and more representative features, could aid in better flexibility and learnability in selecting filters to be pruned without affecting model performance.

\subsection{Pruning in Federated Learning}

Recent advances in pruning in FL have introduced methods to balance model efficiency with collaborative learning. PruneFL~\cite{ref_prunefl} uses a two‐stage process: centralized pruning at a high‐resource client, followed by distributed pruning with federated training guided by gradient importance scores. However, PruneFL ignores client‐specific characteristics, making it unsuitable for non‐IID settings where data distributions vary across clients.
AutoFLIP~\cite{ref_autoflip}, to address client heterogeneity, incorporates federated loss‐exploration to adaptively prune parameters based on client gradient behavior. While it better handles client characteristics than PruneFL, it focuses on a single global model rather than personalized local models and introduces significant computational overhead.
FedMef~\cite{ref_fedmef} emphasizes memory efficiency using scaled activation pruning via NSConv layers, instead of standard convolutional layers, but may blur lesion boundaries in segmentation tasks, thus reducing the segmentation performance. FedDUAP~\cite{ref_fedduap} performs layer‐adaptive pruning based on importance but requires central server-side data access, which may not be useful under strict data privacy regulations. FedDST~\cite{ref_feddst} uses sparse mask votes to construct a global sparse network, yet still prioritizing global sparsity over client‐specific adaptations.

While some studies have explored personalized model pruning~\cite{ref_fedperstrucprune,ref_fedperprune,ref_fedadapprune}, they are limited to unstructured pruning via weight‐masking rather than removing entire components. Also, use of activation features in addition to gradient-based scores as an integrated framework has also not been explored yet. Most FL pruning approaches lack evaluation on segmentation tasks requiring precise boundary detection and introduce impractical computational overheads for real‐world healthcare deployments with constrained resources. In this work, our approach addresses these shortcomings by introducing a two stage, structured pruning framework for federated medical image segmentation that reduces computational overhead, and employs activation statistics-aware pruning to identify redundant structures in a feature-informed manner.

\section{Methodology}
\begin{figure*}
    \centering
    \includegraphics[width=0.9\linewidth]{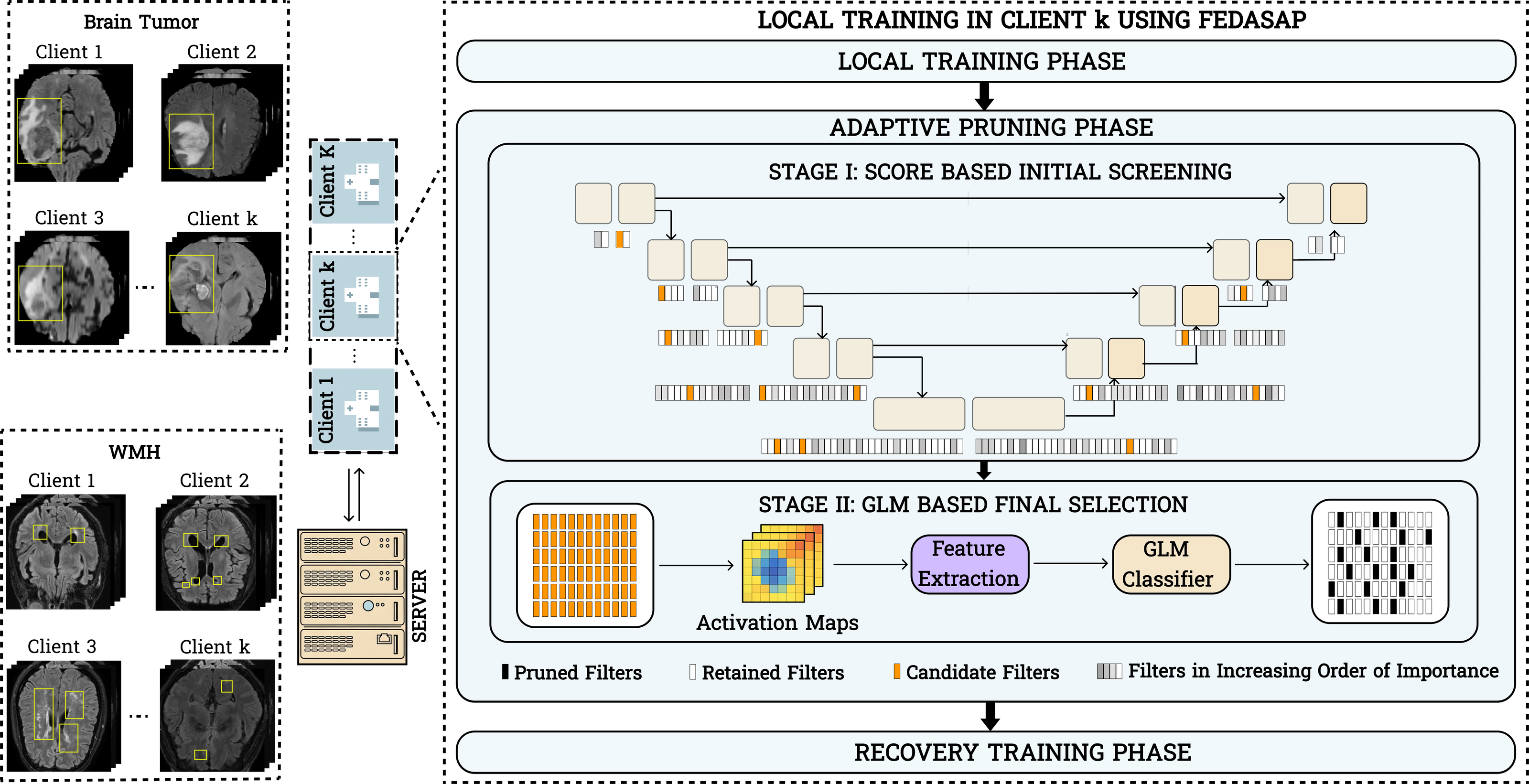} 
    \caption{Overview of the FedASAP framework showing the three-part process: (1) local training (see Sec.~\ref{sssec:train_prune}), (2) adaptive pruning (see Secs.~\ref{sub_adapt},~\ref{adapt_gbfs}), and (3) recovery training (see Sec.~\ref{sssec:rec_prune}). Orange candidate filters have their activations extracted and passed through a GLM classifier for pruning decisions. Filter states indicate pruning history (gray-to-white: increasing importance of unpruned filters, black: ultimately pruned filters, white: ultimately retained filters).}
    \label{fig:fedasap_fig}
\end{figure*}

We propose a two‐stage pruning pipeline, the first score-based screening and the second GLM-based final selection using multiple activation statistics, as illustrated in Figure~\ref{fig:fedasap_fig}. In this section, we will first explain the formulation of the FedASAP setup and the training flow. We will then explain the two-stage pruning framework. 

\subsection{Federated Activation-Statistics Driven Adaptive Pruning (FedASAP)}
\textbf{Preliminaries:} Let $k = {1, 2, \dots, K}$ denote the set of clients. Each client $k \in K$ has a local dataset $\mathcal{D}_k = (X_{img_{k,j}}, Y_{true_{k,j}})_{j=1}^{N_k}$, consisting of $N_k$ input images $X_{img_{k,j}}$ with corresponding segmentation masks $Y_{true_{k,j}}$. Let $\theta_k$ represent the segmentation model at each client $k$. For each training round $r$, during the local training phase over $N$ epochs, $\theta_k$ is trained by minimizing the combination of Dice loss and focal loss~\cite{ref_focal} as follows:
\begin{equation}
    \mathcal{L}_k(\theta_k) = \mathcal{L}_{\mathrm{Focal}} + \frac{1}{N_k}\sum_{j=1}^{N_k}\mathcal{L}_{\mathrm{Dice}}\big(\theta_k(X_{img_{k,j}}),Y_{true_{k,j}}\big)
\end{equation}
where the focal loss is defined as,
\begin{equation}
\mathcal{L}_{\mathrm{Focal}} = -\frac{1}{N_k}\sum_{j=1}^{N_k}(1-\mathrm{Pr}_{k,j})^{\gamma}\,\log(\mathrm{Pr}_{k,j})
\end{equation}
where \(\gamma\) is a focusing parameter and $\mathrm{Pr}_{k,j}$ denotes the predicted output after the sigmoid activation.

The data at each client often has distinct distributions (e.g. different demography, different scanners), resulting in a non-IID setting.
This non-IID nature of the data makes it challenging to train a single generalized global model that work well across all clients. Instead, we aim to train personalized lightweight models $\theta_k$ for each client $k$, tailored to their specific domain and dataset $\mathcal{D}_k$, while leveraging shared knowledge accumulated during the FL process. 

\subsubsection{Training flow with pruning}
\label{sssec:train_prune}
Both the server and each client $k$ maintain a synchronized set $F_k$, containing identifiers (i.e., indices) of all filters pruned in previous rounds (in the initial round, $F_k \in \emptyset$).

\textit{Server-side adaptation: }At each training round $r$, before sending the updated global model to each client, the server derives a client-specific, personalized architecture by applying the accumulated pruning decisions stored in $F_k$.
This reduction step is performed to lower communication cost by transmitting only the relevant, pruned filter indices instead of the full updated global model. The process involves structurally adapting the model by removing filters specified in $F_k$ and eliminating their dependent parameters, such as input channel weights connected to other layers and the corresponding normalization entries, to maintain architectural consistency. Finally, the server sends the personalized reduced model $\theta^r_k$ directly to client $k$.

\textit{Client-side adaptation: }Each client trains for a total of $N$ epochs per round. After receiving the personalized model and completing the standard training for the initial $N/2$ epochs, each client applies the proposed two-stage adaptive pruning approach, detailed in~\ref{sub_adapt} and~\ref{adapt_gbfs} to identify and remove additional unimportant filters. Once filters are selected for removal through this adaptive mechanism, they are appended to the set $F_k$ to be pruned from the network structure in the upcoming rounds and the model is pruned to obtain $\theta_k^{r, pruned}$.

\subsubsection{Recovery training post pruning}
\label{sssec:rec_prune}
Following the pruning, we perform recovery training in which the model is fine-tuned for the latter half of epochs (from $N/2 + 1$). This recovery phase helps the FL framework adapt to the reduced structure (by updating the weights of the remaining filters) and recovers any performance lost due to filter removal.

\subsubsection{Model aggregation}
\label{sssec:model_agg}

After each round $r$, each client sends its updated pruned model (after recovery training) $\theta_k^{r, pruned}$ and the updated set $F_k$ (indices of filters to be pruned) to the server. The server then performs the following operations for each client with $F_k$: firstly, it reconstructs $\theta_k^{r, pruned}$ back to original architecture by zero-masking the pruned parameters so that parameter indices remain consistent. Note that this zero-masking step is done only during aggregation. This step provides a practical workaround that preserves layer index alignment during aggregation for structured pruning (to avoid inconsistencies caused by the removal of different filters/layers in each model). Secondly, the server aligns and aggregates the reconstructed models ${\theta_k^{r, pruned}}_{k \in \mathcal{K}}$ using standard weighted averaging (of non-masked weights) to produce the next global model $\theta^{r+1}$, which serves as the foundation for the subsequent training round.

\subsection{Adaptive Pruning strategy - Stage I: Score-based Initial Screening (SBIS)}
\label{sub_adapt}
The complete workflow of both stages of the proposed adaptive pruning strategy is detailed in Algorithm~\ref{alg:fedasap_adaptive_pruning}. 
In the first stage, filters are coarsely ranked by importance and the lowest‐ranked fraction is flagged as `to be pruned' candidates as follows: each client first computes the importance score for each filter using the first-order Taylor expansion of the loss function~\cite{ref_molchanov} as given by eqn.~\ref{eqn:imp_score}. 
\begin{equation}
    I_{i, k} = \left| g_{i,k} \cdot a_{i, k} \right|
    \label{eqn:imp_score}
\end{equation}

where, $I_{i,k}$ denotes the importance score of filter $i$ for client $k$, where $a_{i,k}$ and $g_{i,k}$ are its activation map and gradient, respectively; '$\cdot$' indicates element-wise multiplication. The filters with the least scores (and hence, minimal contribution) are then flagged as candidates based on a predefined pruning percentage \( p \). This means that that \( p\% \) of the original number of filters are selected as the candidates for pruning in each round. The percentage $p$ is determined empirically, based on the performance and conservation of parameters (refer to Sec.~\ref{ssec:p_results} for more details).

\begin{algorithm}
  \captionsetup{font=footnotesize} 
  \footnotesize
  \caption{Adaptive Pruning Phase of FedASAP}
  \label{alg:fedasap_adaptive_pruning}
  \SetAlgoLined
  \DontPrintSemicolon

  \SetKwInput{KwInput}{Input}
  \SetKwInput{KwOutput}{Output}
  \KwInput{Total $R$ rounds (with warm-up rounds $R_{\text{warm}}$) and $N$ epochs per round, Global model $\theta^r$, local models $\theta_k^r$ trained for first $N$/2 epochs for $K$ clients, GLM $\mathcal{G}_{k}^{(r)}$, pruning \% $p$, threshold $\tau$}
  \KwOutput{Set of filter indices $F_k$, pruned  $\theta_k^{r,\mathrm{pruned}}$ to undergo recovery training for the next $N$/2 epochs until $N^{\text{th}}$ epoch, trained GLM $\mathcal{G}_{k}^{(r)}$}

  \hrule
  Initialize $\theta^0$, $\theta_k^0$,  $\mathcal{D}_{train} \gets \emptyset,  F_k \in \emptyset$\  \\
  \For{round $r=0,\dots,R-1$}{
    \For{client $k=0,\dots,K-1$}{
      \If{$r < R_{\mathrm{warm}}$}{
         \Return \( F_k,\,\theta_k^r\) \hfill \hfill \hfill \hfill \hfill $\blacktriangleright$ No pruning, no SBIS/GBFS\;
      }
      \If{$r = R_{\mathrm{warm}}$}{
        Initialize GLM $\mathcal{G}_{k}^{(r)}$ \hfill \hfill \hfill \hfill \hfill $\blacktriangleright$ Start GLM training\;
        $\theta_k^{r,\mathrm{pruned}} \gets \theta_k^r$ \hfill \hfill \hfill \hfill \hfill \hfill $\blacktriangleright$  No pruning, no GBFS \;
      }

 \nonumber \textbf{Score-based initial screening (SBIS)}\\
     \For {each filter $i$}{
        Collect activations $\{a_{i,k}\}$ and gradients $\{g_{i,k}\}$\;
        $I_{i,k} \gets \bigl|g_{i,k}\cdot a_{i,k}\bigr|$ \hfill \hfill \hfill \hfill \(\blacktriangleright\)\, compute imp. score\;
        Compute sample-averaged activation $\bar{\mathbf{A}}_{i,k}$ and extract features: $L_2$ norm, mean, entropy, DoG mean, DoG entropy, skewness, kurtosis as specified in Section.~\ref{sssec:features}
        $[x_{i1}, x_{i2}, x_{i3}, x_{i4}, x_{i5}, x_{i6}, x_{i7}] \gets calculate\_features(\bar{\mathbf{A}}_{i,k})$
        $\mathbf{x}_i\gets [x_{i1},x_{i2},x_{i3}, x_{i4}, x_{i5}, x_{i6}, x_{i7}]^\top$\;}

      Rank filters by $I_{i,k}$; \emph{Candidates} are the bottom $p\%$ marked for pruning\; 

      \For{each $j\in$ Candidates}{
        $y_j\gets 0$\;
        Add $(\mathbf{x}_j,y_j)$ to $D_{\mathrm{train}}$\;
      }
      Sample a small subset of filters not in \emph{Candidates} for $y_j = 1$ class, and add subset $(x_j, y_j)$ to $D_{train}$\;
      $\mathcal{G}_{k}^{(r)} \gets \text{TrainGLM}(D_{\mathrm{train}})$ \hfill \hfill \hfill \hfill $\blacktriangleright$ train GLM for next round\;  
    \nonumber \textbf{GLM-based final selection (GBFS)}\;
    \If{$r$ $>$\, $R_{\text{warm}}$}{
      \For{each $i\in$ Candidates}{
        \If{$\mathcal{G}_{k}^{(r-1)}(\mathbf{x}_i) < \tau$}{
          $F_k \gets F_k \cup \{i\}$ \hfill \hfill \hfill \hfill \hfill \hfill $\blacktriangleright$ apply GLM from r-1 \;
        }
      }
      \(\theta_k^{r,\mathrm{pruned}} \gets \text{prune } F_k \text{ from } \theta_k^r\)\;
    
      }
      $\mathcal{G}_{k}^{(r-1)} \gets \mathcal{G}_{k}^{(r)}$ \hfill \hfill \;
    
      \Return $F_k,\,\theta_k^{r,\mathrm{pruned}},\,\mathcal{G}_{k}^{(r)}$\;
    }
  }
\end{algorithm}
\subsection{Adaptive Pruning strategy - Stage II: GLM-based Final Selection (GBFS)} 
\label{adapt_gbfs}

Pruning with a single heuristic, such as weight magnitude or a first‐order Taylor score, $I_{i,k}$, offers only a limited perspective on filter importance~\cite{ref_blendcriteria}. In practice, combining complementary attributes has the potential to produce more relevant compression. Therefore, in the second stage each candidate filter is passed through a client‐specific GLM fitted on activation statistics, which determines whether the filter should be pruned or retained. This learned decision boundary integrates these diverse attributes, preserving filters exhibiting distinctive activation patterns while eliminating truly redundant ones. 

\subsubsection{Generalized Linear Model formulation}
We chose a GLM model for filter selection due to its computational efficiency and theoretical interpretability. In GLM, the conditional expectation of the binary pruning decision is modelled through a linear predictor transformed by a link function:
\begin{equation}
g\bigl(\mathbb{E}[y \mid \mathbf{x}]\bigr) = \mathbf{x}^\top \boldsymbol{\beta},
\end{equation}
where $g(\cdot)$ represents the logit link function, $y \in \{0,1\}$ denotes the binary pruning decision, $\mathbf{x} \in \mathbb{R}^d$ is the filter feature vector, and $\boldsymbol{\beta} \in \mathbb{R}^d$ represents the learned parameter vector. Filters are pruned when the probabilistic output of the GLM is below the predetermined threshold $\tau$ and vice versa.

\subsubsection{Feature design via multi-component filter profiling}
\label{sssec:features}
For each filter $i$, 
given the filter's activation tensor $\mathbf{A}_i \in \mathbb{R}^{N_k \times 1 \times d \times h \times w}$ across $N_k$ samples, we first compute the sample-wise average, $\bar{\mathbf{A}}_i = \frac{1}{N_k} \sum_{j=1}^{N_k} \mathbf{A}_{i,j}$, yielding averaged activations $\bar{\mathbf{A}}_i \in \mathbb{R}^{d \times h \times w}$. From these sample-averaged activations, we extract the following features:
\paragraph{\textbf{Activation Energy}} The $L_2$-norm serves as a smooth aggregate measure of the filter's output energy:
\begin{equation}
x_{i1} = \|\bar{\mathbf{A}}_i\|_2 = \sqrt{\sum_{u,v,z} \bar{A}_{i,u,v,z}^2}
\end{equation}
A larger $L_2$-norm indicates stronger overall activations and greater influence on the network. The $L_2$-norm provides a stable metric of filter strength that informs the classifier about the scale of a filter's response.
\paragraph{\textbf{Baseline Activation Level}}The mean captures the baseline offset or bias of the filter's output:
\begin{equation}
x_{i2} = \frac{1}{d \times h \times w} \sum_{u,v,z} \bar{A}_{i,u,v,z}
\end{equation}

\paragraph{\textbf{Entropy}} The entropy measures distributional complexity using histogram-based estimation:
\begin{equation}
x_{i3} = -\sum_{c=1}^{B} p_c \log p_c, \quad \text{where } p_c = \frac{h_c}{\sum_{j=1}^{B} h_j}
\end{equation}
Here, $h_c$ represents the histogram count in bin $c$ for $B$ uniformly spaced bins across the range $[\min(\bar{\mathbf{A}}_i), \max(\bar{\mathbf{A}}_i)]$. High entropy indicates diverse activation distributions, suggesting information-rich filter behavior. Entropy helps identify filters whose activations are more informative or complex, with prior work showing that feature maps with higher entropy contain more fine-grained knowledge  ~\cite{ref_filterpruningbasedinformation}.
\paragraph{\textbf{Kurtosis}} The kurtosis quantifies the tailedness or peakness of the filter's output:
\begin{equation}
x_{i4} \;=\; \frac{\dfrac{1}{d \times h \times w}\sum_{u,v,z}\big(\bar{A}_{i,u,v,z}-\mu_i\big)^4}
{\Big(\dfrac{1}{d \times h \times w}\sum_{u,v,z}\big(\bar{A}_{i,u,v,z}-\mu_i\big)^2\Big)^{2}} \;-\; 3
\
\end{equation}
where $\mu_i$ denotes the mean of $\bar{A}_{i,u,v,z}$. A positive value of kurtosis indicates heavy tails or the presence of outlier/high-magnitude activations, while a negative value indicates uniform activation values.
\paragraph{\textbf{Skewness}} The skewness measures the asymmetry of the activation distribution around its mean:
\begin{equation}
x_{i5} \;=\; \frac{\dfrac{1}{d \times h \times w}\sum_{u,v,z}\big(\bar{A}_{i,u,v,z}-\mu_i\big)^3}
{\Big(\sqrt{\dfrac{1}{d \times h \times w}\sum_{u,v,z}\big(\bar{A}_{i,u,v,z}-\mu_i\big)^2}\;\Big)^{3}}
\
\end{equation}
\paragraph{\textbf{Mean of Difference of Gaussian (DoG)}}
Let \(\mathrm{DoG}(\cdot)\) denote the 2D Difference-of-Gaussians operator applied slice-wise to the activation volume \(\bar{\mathbf{A}}_i\). We define the DoG response tensor:
\begin{equation}
    \bar{\mathbf{D}}_i \;=\; \big|\,\mathrm{DoG}(\bar{\mathbf{A}}_i)\,\big|
\end{equation}
The mean of the DoG response for filter \(i\) is calculated same as baseline activation level, and is denoted by $x_{i6}$. This measures the average edge response strength of the filter at the chosen DoG scales.
\paragraph{\textbf{Entropy of DoG}}
The DoG entropy quantifies the spatial variability of the DoG response, and is denoted by $x_{i7}$.

The multi-dimensional feature vector $\mathbf{x_i}=[x_{i1}, x_{i2}, x_{i3}, ..., x_{i7}]^\top$ succinctly captures filter $i$'s behavior, providing a signature of each filter's importance, enabling our framework to prune in a principled way without relying on a single criterion. Fitting the weight vector $\boldsymbol{\beta}$ via convex maximum likelihood in this low‐dimensional space entails negligible computational overhead, enabling each client in the federated setting to adapt its pruning strategy efficiently and robustly, even with limited samples.

\subsubsection{GLM-guided final pruning}
During an initial warm-up period (\( r < R_{\text{warm}} \)), no pruning is applied. At \( r = R_{\text{warm}} \), although we do not begin pruning, we initialize the GLM classifier training \( \mathcal{G}_k^{(r)} \) for the adaptive pruning process of the subsequent round \( r > R_{\text{warm}}\). 

From \( r > R_{\text{warm}}\) onward, the GLM classifier is trained in a similar manner to the temporal ensembling \cite{laine2017temporal}: at each round \( r \), the client applies the pretrained GLM classifier \( \mathcal{G}_k^{(r-1)} \) (from the previous round) to predict among the candidate filters (marked for pruning in SBIS, see Sec.~\ref{sub_adapt}) for final pruning. For the initial round (\( r > R_{\text{warm}}\)),  the feature vector $\mathbf{x_i}$ and the corresponding $I_{i, k}$ values collected from \( r = R_{\text{warm}} \) are used for training the GLM. During inference, the classifier \( \mathcal{G}_k^{(r-1)} \) predicts a ``keep" probability using the feature vector $\mathbf{x_i}$. If the predicted probability is below a predefined retention threshold \( \tau \), the filter is removed along with its dependent channels and normalization parameters, producing the reduced model \( \theta_k^{r,\text{pruned}} \). The threshold \( \tau \) controls pruning aggressiveness: lower values result in more conservative pruning, while higher values lead to more aggressive pruning (refer to Sec.~\ref{ssec:imp_details} for more details on $\tau$). Once the pruning is performed, the client trains \( \mathcal{G}_k^{(r)} \) for the next round. To construct training data for \( \mathcal{G}_k^{(r)} \), $\mathbf{x_i}$ is computed for all candidate filters (prune class, $y=0$) from the current round. To create a contrasting ``keep'' class ($y=1$), a small random subset of non-candidate filters is also sampled, and identical features are computed, yielding a balanced dataset for  training.

\section{Experimental Setup} 

\subsection{Dataset Details}

We evaluate the proposed method on two neuroimaging datasets: MICCAI White Matter Hyperintensities (WMH) challenge dataset~\cite{ref_wmh} and Federated Tumor Segmentation (FeTS) dataset~\cite{ref_fets,ref_openfl,ref_tcga}, consisting of FLAIR MRI volumes. The datasets are multi-centric and heterogeneous, with WMH acquired from 5 different scanners and subset of FeTS from 4 centres. The training/validation/test splits of the individual clients consisting of the above datasets are reported in Table~\ref{tab:dataset-splits}.
The WMH cohort requires the model to prioritize fine‑grained, high‑resolution detail in order to detect numerous small, subtle WMH lesions, whereas the FeTS dataset needs emphasis on coarse structural context to accurately segment large, irregularly shaped tumors. Additionally, each client consists of just 3–12 training examples, representing the privacy‑sensitive, low-data regimes often found in real‑world federated deployments. Furthermore, the data sets pose substantial non‑IID heterogeneity arising from disparate imaging protocols, scanner hardware, and patient demographics, as well as variability in lesion burden.

\begin{table*}[ht]
\centering
\caption{\textsc{Dataset Splits (Train / Validation / Test) for WMH and FeTS.}}
\label{tab:dataset-splits}
\begingroup
\footnotesize          
\setlength{\tabcolsep}{4pt}
\renewcommand{\arraystretch}{0.95}
\begin{tabular}{llccc|llccc}
\toprule
\textbf{Dataset} & \textbf{Site} & \textbf{Train} & \textbf{Val} & \textbf{Test}
 & \textbf{Dataset} & \textbf{Site} & \textbf{Train} & \textbf{Val} & \textbf{Test} \\
\midrule
\multirow{5}{*}{WMH} & 1 & 12 & 5 & 15 & \multirow{4}{*}{FeTS} & 1 & 12 & 8 & 10 \\
                     & 2 & 12 & 5 & 15 &                       & 2 & 8  & 5 & 10 \\
                     & 3 & 12 & 5 & 15 &                       & 3 & 3  & 1 & 10 \\
                     & 4 & 6  & 1 & 3  &                       & 4 & 9  & 3 & 10 \\
                     & 5 & 6  & 1 & 3  &                       &        &    &   &    \\
\bottomrule
\end{tabular}
\endgroup
\end{table*}

\noindent {\textbf{Evaluation metrics.}} We use Dice Similarity Coefficient (DSC) and lesion-level F1 score (LL-F1) to evaluate segmentation performance, and track parameter count (PC) for pruning effectiveness, indicating model compactness. Additionally, we perform a Wilcoxon signed-rank test to evaluate the statistical significance of the differences of ablation results/FedASASP versus FedAvg+SBIS for the 4-layer U-Net and deeper architectures (significant at $p<0.05$).

\subsection{Implementation Details}
\label{ssec:imp_details}
Experiments are conducted on a Linux workstation with an Intel i9-10980XE CPU, 128 GB RAM, and two NVIDIA RTX A6000 GPUs (each with 48 GB memory). For $\theta_k^r$ in each client $k$, we use the U-Net architecture  comprising 4 encoder and decoder blocks, with the initial number of features set to 16. Local training in each client is performed using the Adam optimizer (initial learning rate = $0.0005$, $\epsilon = 10^{-4}$). Data augmentation includes random rotation ($[-10 ^\circ,10 ^\circ]$), horizontal flip, Gaussian noise injection ($\sigma \in [0.1, 0.2]$) and intensity rescaling. We train for 100 rounds, with empirically chosen parameters: batch size of $4$, with $N = 10$ and pruning percentage $p = 1$. We define the number of warm-up rounds $R_{warm}=10$, and sample 5\% of the non-candidate filters to create the ``keep'' class mentioned in~\ref{sub_adapt}. In feature extraction, DoG filters were applied on activation maps at two scales (with $\sigma_1 = 0.2$ and $\sigma_2 = 0.5$) and for GLM prediction, we use $\tau=0.8$ for WMH and $\tau=0.9$ for FeTS.

\subsection{Experiments}
\label{ssec:exps}
\subsubsection{Comparison with the state-of-the-art methods}
\label{sssec:sota_comp}
We compare FedASAP with the state-of-the-art FL methods for non-IID settings, such as FedProx \cite{ref_fedprox}, FedBN \cite{ref_fedbn}, and MOON \cite{ref_moon}, alongside standard FedAvg \cite{ref_fedavg} and centralized training. 
Additionally, we further compare with the recent federated pruning approaches: PruneFL \cite{ref_prunefl}, AutoFLIP \cite{ref_autoflip}, and FedDST \cite{ref_feddst}. Some prior methods could not be included for comparison due to constraints (e.g., the need for server-side data access \cite{ref_fedduap}). Since the methods used for comparison require target sparsity to be specified at initialization rather than using incremental pruning, we set the target sparsity to match that achieved by our best-performing setting in FedASAP: $73.0\%$ for WMH and $45.6\%$ for FeTS.
For AutoFLIP's exploration stage, we trained models at each client for $100$ epochs. For FedDST and PruneFL, pruning masks were readjusted every five rounds. All other parameters were kept consistent with those specified in Sec.~\ref{ssec:imp_details}.
For FedDST, for the WMH dataset, we doubled $N$ to be able to achieve representative performance, incurring extra computational expense.

\noindent {\textbf{Communication Overhead Analysis.}} We also analyze the communication overhead (beyond standard model weight updates) of the above state-of-the-art pruning methods relative to FedASAP, assuming 32-bit floating-point representation.

\subsubsection{Ablation study: Effect of two-stage adaptive pruning in FedASAP}
We study the effect of the adaptive pruning by comparing FedASAP with, firstly, the FedAvg + SBIS modules. Secondly, to further study the effectiveness of GLM, we replace GLM-based final pruning with a simple, lesion‐informed, layer‐adaptive thresholding (LAT) strategy (explained below). 

\noindent {\textbf{Layer-adaptive thresholding (LAT) strategy.}} 
In LAT strategy, in the two-stage approach to pruning, after SBIS, instead of using GLM, we introduce an adaptive thresholding approach. We use the idea that lesion size naturally correlates with feature granularity--small lesions rely on fine textures captured by early encoder and late decoder layers, while large lesions depend on broader structural cues encoded in deeper layers. So, tailoring the extent of pruning in specific layers based on lesion size can lead to client-specific lesion segmentation. To achieve this, after selecting candidates by applying $p$\% in SBIS, we assign each layer $\ell$ a retention threshold $\tau^{\ell}_k \in (0,1)$. For every candidate filter $j$, we draw a random value $q \in (0,1)$. If $q > \tau^{\ell(j)}_k$ for layer $\ell(j)$, the filter is retained, otherwise it is pruned. For \emph{small lesions}, lower $\tau^{\ell}_k$ in early encoder and late decoder layers (shallow layers) preserves fine‐grained features. In contrast, for \emph{large lesions}, lower $\tau^{\ell}_k$ in deep encoder and early decoder layers (deeper layers) preserves coarse, structural features. Based on this, we consider two settings in LAT (chosen empirically):

\noindent \textit{Positive LAT (PLAT):} $\tau^{\ell}_k$ ramps linearly from 0.3 at the first encoder block to 0.7 at the bottleneck, then back to 0.3 at the final decoder layer, thereby retaining more filters in shallow layers which related to finer features.

\noindent \textit{Negative LAT (NLAT):} $\tau^{\ell}_k$ descends from 0.7 at the encoder to 0.3 at the bottleneck, then ascends to 0.7 at the decoder, emphasizing the retention of coarser, structural information.

\subsubsection{Effect of pruning on computational efficiency} 
\label{sssec:abl_study1}
We evaluate the impact of explicit structural pruning versus the unpruned baseline (including dense/unstructured sparse full-sized architectures) on resource usage. Specifically, we measure (i)   FLOPs per forward pass, (ii) inference time per sample, and (iii) peak memory usage during inference for both WMH and FeTS datasets. We measure the above on CPU for a realistic comparison to reflect resource-constrained deployment scenarios where GPU inference is not possible.

\subsubsection{Extensibility to deeper architectures} To demonstrate the extensibility of the FedASAP adaptive pruning approach beyond the standard 4-layer U-Net in FL settings, we adapted FedASAP (including SBIS and GBFS) to a deeper 5-layer U-Net backbone. 

\subsubsection{Effect of the pruning parameter p}
We study the influence of the SBIS pruning percentage $\textit{p}$, which controls the degree of structural sparsity during training. Specifically, we vary $\textit{p}$ across a range of values [0.5, 1, 2] and measure its impact on (i) model size reduction over training and (ii) segmentation accuracy for both WMH and FeTS clients.

\section{Results}
\label{sec:results}

\subsection{Comparison with the state-of-the-art methods} 
\label{ssec:sota_comp_res}
\begin{figure*}
    \centering
    \includegraphics[width=0.9\linewidth]{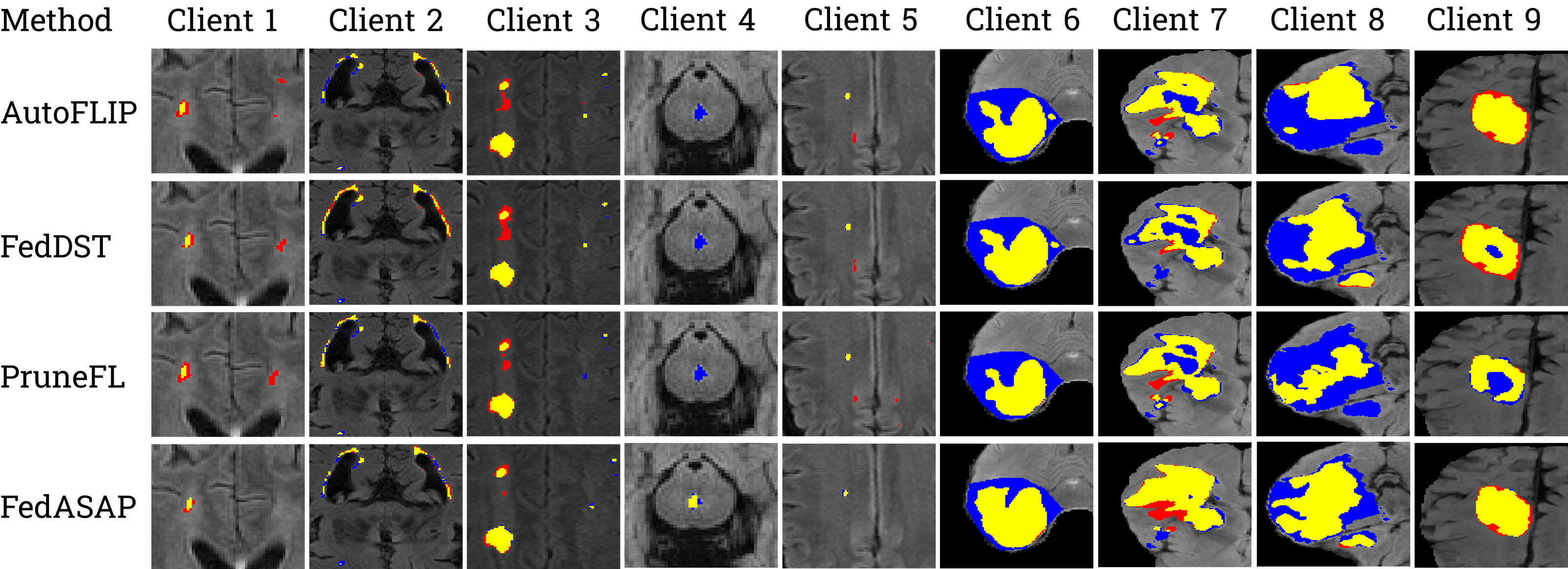} 
    \caption{Comparison of  FedASAP with the state-of-the-art FL pruning approaches on the WMH and FeTS datasets. True positive, false positive and false negatives voxels are indicated in yellow, red and blue, respectively.}
    \label{fig:vis_res}
\end{figure*}

Table~\ref{tab:wmh_fets_metrics} reports the results of the comparison of FedASAP with state-of-the-art methods, and Figure~\ref{fig:vis_res} shows the visual results. On the WMH dataset, FedASAP achieves the best Dice score (0.743), outperforming FedAvg by 0.039, and the highest LL-F1 score (0.591) while reducing the network parameters by 73.0\% (from 5.6M to 1.5M). On the FeTS dataset, our approach achieves the best Dice score (0.796), surpassing all the state-of-the-art methods but the centralized training scenario, while maintaining a competitive LL-F1 score of 0.660 with 45.6\% reduction in parameters (5.6M to 3.0M). For WMH and FeTS, FedASAP reduces false positives and detects more true positive lesion voxels as shown in Figure~\ref{fig:vis_res}.

FedASAP also consistently outperforms all unstructured pruning approaches. For instance, on the WMH dataset, FedASAP achieves a Dice score of 0.743, surpassing FedDST (0.741), AutoFLIP (0.735), and PruneFL (0.721) by 0.3\%, 1.1\%, and 3.1\%, respectively. On the FeTS dataset, the performance gap is more pronounced: FedASAP (0.796) outperforms FedDST (0.782), AutoFLIP (0.780), and PruneFL (0.766) by 1.8\%, 2.1\%, and 3.9\%, respectively.
Table~\ref{tab:comm_compare} shows the comparison of additional communication overhead of various federated pruning methods, and exact values per client for a 4-layer deep U-Net (\(n=5.6\times10^6\) parameters, \(N_F=1456\) filters, \(T=100\) rounds, \(\Delta R=5\)). In each case, the additional bits arise as follows: 
\textbf{PruneFL:} Full 32‑bit gradient vectors (one per parameter) are exchanged every \(\Delta R\) rounds, and the updated binary mask (one bit per parameter) is returned at the same frequency.
\textbf{FedDST:} Only binary masks (one bit per parameter) are sent and received every \(\Delta R\) rounds—no gradients are exchanged.
\textbf{AutoFLIP:} A one‑time float32 guidance matrix (one 32‑bit value per parameter) is uploaded before pruning, and then a single binary mask (one bit per parameter) is downloaded after pruning.
\textbf{FedASAP:} Clients transmit only the indices of pruned filters (each encoded in \(\lceil\log_2 N_F\rceil\) bits) once per filter; there is no extra download beyond standard model weight updates. Hence, FedASAP transmits the lowest with an upload of 2 KB (reducing the overhead by the factor of $10^8$ compared to other methods) with no extra download.

\subsection{Ablation study: Effect of two-stage adaptive pruning in FedASAP}
Table \ref{tab:wmh_fets_metrics} (lower panel) shows the ablation results of the effect of individual steps in adaptive pruning. Using FedAvg + SBIS achieve the lightest network (0.8M parameters) at the cost of the lowest DSC and LL-F1 values. With the addition of LAT module, for WMH segmentation—dominated by small lesions—PLAT yields an 8.9\% ($p<0.001$) improvement in DSC and a 21.6\% ($p<0.001$) gain in LL‑F1 over FedAvg + SBIS. Although NLAT achieves similar segmentation performance (DSC of 0.722 vs. 0.732), it results in a much larger network (2.9M vs. 1.3M parameters). In contrast, NLAT outperforms its positive counterpart by 6.2\% Dice (0.787 vs 0.741) on the larger lesions in FeTS. FedASAP’s use of GLM-based refinement of pruning decision further boosts performance. On WMH, it raises Dice by approximately 10.5\% ($p<0.001$) over SBIS (0.743 vs 0.672), and on FeTS, by over 6.8\%($p<0.001$) (0.796 vs 0.745). Moreover, compared to the best ablation variant, FedASAP adds 1.1\% Dice and 9.3\% LL‑F1 on FeTS and 1.5\% Dice and 2.8\% LL‑F1 on WMH. 

\begin{table*}
\centering
\caption{Comparison of Segmentation Performance on WMH and FeTS Datasets. Mean Values Across Clients Reported ; Best Performance Among FL Methods Shown in Bold. ** Indicates $p<0.005$, Representing a Significant Improvement Over FedAvg+SBIS (PC: Parameter Count).}
\label{tab:wmh_fets_metrics}
\setlength{\tabcolsep}{15pt} 
\renewcommand{\arraystretch}{1} 
\begin{threeparttable}
\begin{tabular}{@{} l @{\hskip 12pt} c @{\hskip 12pt} c @{\hskip 12pt} c @{}}
\toprule
\textbf{Method} & \makecell{\textbf{DSC} \\ \textbf{Mean [95\% CI]}} & \makecell{\textbf{LL-F1} \\ \textbf{Mean [95\% CI]}} & \makecell{\textbf{PC (M)} \\ \textbf{(\% remaining)}} \\
\midrule
\multicolumn{4}{c}{\textbf{WMH Dataset}} \\
\midrule
Centralized Training & $0.731\hspace{0.5em}\text{\tiny[0.695,\,0.767]}$ & $0.571\hspace{0.5em}\text{\tiny[0.535,\,0.607]}$ & $5.6$ ($100\%$) \\
FedProx ($\lambda = 0.04$) \cite{ref_fedprox} & $0.715\hspace{0.5em}\text{\tiny[0.681,\,0.749]}$ & $0.590\hspace{0.5em}\text{\tiny[0.559,\,0.621]}$ & $5.6$ ($100\%$) \\
FedBN \cite{ref_fedbn} & $0.707\hspace{0.5em}\text{\tiny[0.668,\,0.746]}$ & $0.570\hspace{0.5em}\text{\tiny[0.539,\,0.601]}$ & $5.6$ ($100\%$) \\
MOON ($\mu = 0.1$) \cite{ref_moon} & $0.716\hspace{0.5em}\text{\tiny[0.682,\,0.75]}$ & $0.587\hspace{0.5em}\text{\tiny[0.557,\,0.617]}$ & $5.6$ ($100\%$) \\
FedAvg \cite{ref_fedavg} & $0.714\hspace{0.5em}\text{\tiny[0.675,\,0.753]}$ & $0.584\hspace{0.5em}\text{\tiny[0.553,\,0.615]}$ & $5.6$ ($100\%$) \\
PruneFL \cite{ref_prunefl} & $0.721\hspace{0.5em}\text{\tiny[0.682,\,0.76]}$ & $0.556\hspace{0.5em}\text{\tiny[0.517,\,0.595]}$ & $1.5$ ($27.0\%$) \\
AutoFLIP \cite{ref_autoflip} & $0.735\hspace{0.5em}\text{\tiny[0.700,\,0.77]}$ & $0.563\hspace{0.5em}\text{\tiny[0.527,\,0.599]}$ & $1.5$ ($27.0\%$) \\
FedDST \cite{ref_feddst} & $0.741\hspace{0.5em}\text{\tiny[0.707,\,0.775]}$ & $0.589\hspace{0.5em}\text{\tiny[0.555,\,0.623]}$ & $1.5$ ($27.0\%$) \\
FedAvg + SBIS & $0.672\hspace{0.5em}\text{\tiny[0.634,\,0.710]}$ & $0.473\hspace{0.5em}\text{\tiny[0.436,\,0.510]}$ & $0.8$ ($14.3\%$) \\
FedAvg + SBIS + PLAT & $0.732^{**}\hspace{0.5em}\text{\tiny[0.696,\,0.768]}$ & $0.575^{**}\hspace{0.5em}\text{\tiny[0.539,\,0.611]}$ & $1.3$ ($23.2\%$) \\
FedAvg + SBIS + NLAT & $0.722\hspace{0.5em}\text{\tiny[0.686,\,0.758]}$ & $0.591\hspace{0.5em}\text{\tiny[0.555,\,0.627]}$ & $2.9$ ($51.8\%$) \\
FedASAP ($\tau = 0.8/0.9$) & $\mathbf{0.743^{**}}\hspace{0.5em}\text{\textbf{\tiny[0.707,\,0.779]}}$ & $\mathbf{0.591^{**}}\hspace{0.5em}\text{\textbf{\tiny[0.554,\,0.628]}}$ & $\mathbf{1.5}$ ($27.0\%$) \\
\midrule
\multicolumn{4}{c}{\textbf{FeTS Dataset}} \\
\midrule
Centralized Training & $0.804\hspace{0.5em}\text{\tiny[0.755,\,0.853]}$ & $0.610\hspace{0.5em}\text{\tiny[0.517,\,0.703]}$ & $5.6$ ($100\%$) \\
FedProx ($\lambda = 0.04$) \cite{ref_fedprox} & $0.784\hspace{0.5em}\text{\tiny[0.725,\,0.843]}$ & $0.570\hspace{0.5em}\text{\tiny[0.476,\,0.664]}$ & $5.6$ ($100\%$) \\
FedBN \cite{ref_fedbn} & $0.785\hspace{0.5em}\text{\tiny[0.724,\,0.846]}$ & $0.541\hspace{0.5em}\text{\tiny[0.434,\,0.648]}$ & $5.6$ ($100\%$) \\
MOON ($\mu = 0.1$) \cite{ref_moon} & $0.779\hspace{0.5em}\text{\tiny[0.708,\,0.85]}$ & $0.605\hspace{0.5em}\text{\tiny[0.503,\,0.707]}$ & $5.6$ ($100\%$) \\
FedAvg \cite{ref_fedavg} & $0.784\hspace{0.5em}\text{\tiny[0.700,\,0.868]}$ & $0.555\hspace{0.5em}\text{\tiny[0.462,\,0.648]}$ & $5.6$ ($100\%$) \\
PruneFL \cite{ref_prunefl} & $0.766\hspace{0.5em}\text{\tiny[0.704,\,0.828]}$ & $0.590\hspace{0.5em}\text{\tiny[0.494,\,0.686]}$ & $3.0$ ($54.4\%$) \\
AutoFLIP \cite{ref_autoflip} & $0.780\hspace{0.5em}\text{\tiny[0.715,\,0.845]}$ & $0.604\hspace{0.5em}\text{\tiny[0.505,\,0.703]}$ & $3.0$ ($54.4\%$) \\
FedDST \cite{ref_feddst} & $0.782\hspace{0.5em}\text{\tiny[0.720,\,0.844]}$ & $0.634\hspace{0.5em}\text{\tiny[0.541,\,0.727]}$ & $3.0$ ($54.4\%$) \\
FedAvg + SBIS & $0.745\hspace{0.5em}\text{\tiny[0.696,\,0.794]}$ & $0.647\hspace{0.5em}\text{\tiny[0.607,\,0.687]}$ & $2.2$ ($39.3\%$) \\
FedAvg + SBIS + PLAT & $0.741\hspace{0.5em}\text{\tiny[0.686,\,0.796]}$ & $0.609\hspace{0.5em}\text{\tiny[0.517,\,0.701]}$ & $2.3$ ($41.1\%$) \\
FedAvg + SBIS + NLAT & $0.787^{**}\hspace{0.5em}\text{\tiny[0.738,\,0.836]}$ & $0.580\hspace{0.5em}\text{\tiny[0.488,\,0.672]}$ & $3.6$ ($64.3\%$) \\
FedASAP ($\tau = 0.8/0.9$) & $\mathbf{0.796^{**}}\hspace{0.5em}\text{\textbf{\tiny[0.747,\,0.845]}}$ & $\mathbf{0.660^{**}}\hspace{0.5em}\text{\textbf{\tiny[0.564,\,0.756]}}$ & $\mathbf{3.0}$ ($54.4\%$) \\
\bottomrule
\end{tabular}
\begin{tablenotes}[flushleft]
\setlength\labelsep{0pt}
\linespread{1}\small
\item For fair comparison, the best-performing local model for each client was selected, rather than using a common global model. FedASAP used \( \tau = 0.8 \) for WMH and \( \tau = 0.9 \) for FeTS.
\end{tablenotes}
\end{threeparttable}
\end{table*}

\begin{table*}
\footnotesize
\setlength{\tabcolsep}{0pt}
\begin{threeparttable}
\caption{Communication Overhead Per Client for a 4-Layer Deep U-Net.}
\label{tab:comm_compare}
\begin{tabular*}{\linewidth}{@{\extracolsep{\fill}}lcc} 
\toprule
\textbf{Method}    & \textbf{Upload} & \textbf{Download} \\
\midrule
PruneFL   & $\mathbf{32nT/\Delta R}$ \textbf{bits} & $\mathbf{nT/\Delta R}$ \textbf{bits} \\
& $3.584\times10^9$ bits ($448$ MB) & $1.12\times10^8$ bits ($14$ MB) \\
FedDST    & $\mathbf{nT/\Delta R}$ \textbf{bits} & $\mathbf{nT/\Delta R}$ \textbf{bits} \\
& $1.12\times10^8$ bits ($14$ MB)   & $1.12\times10^8$ bits ($14$ MB) \\
AutoFLIP  & $\mathbf{32n}$ \textbf{bits (once)} & $\mathbf{n}$ \textbf{bits (once)} \\
& $1.792\times10^8$ bits ($22.4$ MB) & $5.6\times10^6$ bits ($0.7$ MB) \\
FedASAP   & $\mathbf{N_F\lceil\log_2 N_F\rceil}$ \textbf{bits (max)} & \textbf{0 bits} \\
& $16{,}016$ bits ($2$ KB) & 0 bits \\ 
\bottomrule
\end{tabular*}
\begin{tablenotes}[flushleft]
\setlength\labelsep{0pt} 
\linespread{1}\small
\item $n$ = no. of parameters, $N_F$ = no. of filters, $T$ = total no. of rounds, $\Delta R$ = time period between mask readjustments (in rounds).
\end{tablenotes}
\end{threeparttable}
\end{table*}
\subsection{Effect of pruning on computational efficiency} 
\label{ssec:p_results}
\begin{figure*}
    \centering
    \includegraphics[width=13cm]{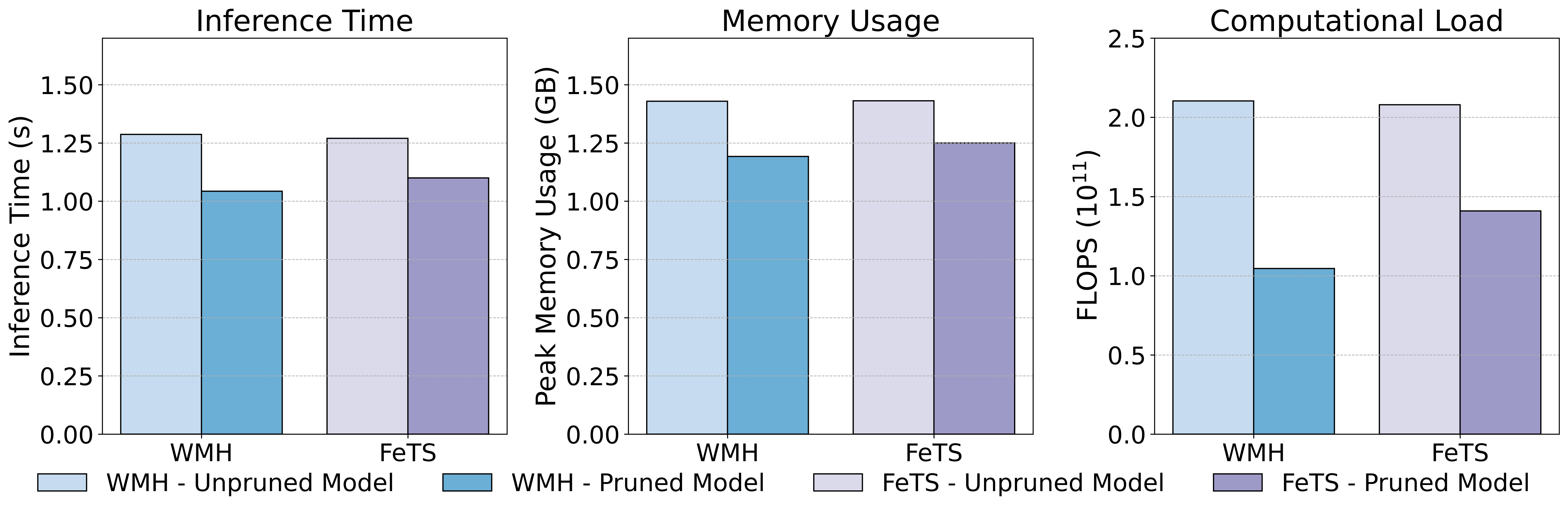}  
    \caption{Comparison of inference time, peak memory usage and computational load between the unpruned and pruned models across WMH and FeTS datasets. The unpruned model denotes dense and sparse full-sized architectures that do not offer computational savings. The pruned Model represents architectures reduced via explicit structural pruning. All values are reported per forward pass, averaged over test samples, and across clients.}
    \vspace{-1em}
    \label{fig:metrics-comparison}
\end{figure*}

As shown in Figure~\ref{fig:metrics-comparison}, for the WMH dataset, pruning reduces the computational load by 50\%, from \(2.104\times10^{11}\)FLOPs to \(1.046\times10^{11}\)FLOPs per forward pass, while for the FeTS dataset, the reduction is 32.2\%. Inference time per sample likewise decreases by 19.0\% (from 1.287s to 1.043s) on WMH clients and 13.4\% (from 1.270s to 1.100s) on FeTS clients. Peak memory usage during inference falls by 16.6\% (1.429 GB\(\rightarrow\)1.192 GB) on WMH and 12.6\% (1.431 GB\(\rightarrow\)1.250 GB) on FeTS.
\subsection{Extensibility to deeper architectures} Table~\ref{tab:deep_unet_results} summarizes the results. The deeper U-Net is over-parameterized in the case of FedAvg for WMH and attains a DSC of 0.686. Applying SBIS alone improves WMH performance to 0.711 (+3.6\% relative to FedAvg), and addition of the subsequent GBFS stage further increases DSC and LL-F1 to 0.746 and 0.603 respectively (+4.9\% and +7.5\% relative to SBIS, both improvements statistically significant with $p<0.001$) while pruning more weights (73.0\% after GBFS vs.\ 69.3\% after SBIS). On FeTS, SBIS causes a small initial drop (–0.9\%), but the GBFS stage recovers and ultimately outperforms FedAvg by 2.0\% in DSC and 5.3\% in LL-F1, with 58.6\% of parameters pruned. 

\begin{table*}
\centering
\caption{Segmentation Performance with a Deeper U-Net Backbone (5 Encoder–Decoder Blocks). Mean Values Across Clients Reported (Best Performance Shown in Bold). ** Indicates $p<0.005$, Representing a Significant Improvement Over FedAvg+SBIS (PC: Parameter Count).}
\label{tab:deep_unet_results}
\setlength{\tabcolsep}{15pt} 
\renewcommand{\arraystretch}{1} 
\begin{threeparttable}
\begin{tabular}{@{} l @{\hskip 12pt} c @{\hskip 12pt} c @{\hskip 12pt} c @{}}
\toprule
\textbf{Method} & \makecell{\textbf{DSC} \\ \textbf{Mean [95\% CI]}} & \makecell{\textbf{LL-F1} \\ \textbf{Mean [95\% CI]}} & \makecell{\textbf{PC (M)} \\ \textbf{(\% remaining)}} \\
\midrule
\multicolumn{4}{c}{\textbf{WMH Dataset}} \\
\midrule
FedAvg \cite{ref_fedavg} & $0.686\hspace{0.5em}\text{\tiny[0.643,\,0.729]}$ & $0.599\hspace{0.5em}\text{\tiny[0.569,\,0.629]}$ & $22.6$ ($100\%$) \\
FedAvg + SBIS & $0.711\hspace{0.5em}\text{\tiny[0.676,\,0.746]}$ & $0.561\hspace{0.5em}\text{\tiny[0.524,\,0.598]}$ & $6.9$ ($30.7\%$) \\
FedASAP ($\tau = 0.8$) & $\mathbf{0.746^{**}}\hspace{0.5em}\text{\textbf{\tiny[0.711,\,0.781]}}$ & $\mathbf{0.603^{**}}\hspace{0.5em}\text{\textbf{\tiny[0.570,\,0.636]}}$ & $\mathbf{6.1}$ ($27.0\%$) \\
\midrule
\multicolumn{4}{c}{\textbf{FeTS Dataset}} \\
\midrule
FedAvg \cite{ref_fedavg} & $0.794\hspace{0.5em}\text{\tiny[0.730,\,0.858]}$ & $0.585\hspace{0.5em}\text{\tiny[0.505,\,0.665]}$ & $22.6$ ($100\%$) \\
FedAvg + SBIS & $0.787\hspace{0.5em}\text{\tiny[0.728,\,0.846]}$ & $0.610\hspace{0.5em}\text{\tiny[0.528,\,0.692]}$ & $11.9$ ($52.5\%$) \\
FedASAP ($\tau = 0.9$) & $\mathbf{0.810}\hspace{0.5em}\text{\textbf{\tiny[0.758,\,0.862]}}$ & $\mathbf{0.616\hspace{0.5em}\text{\textbf{\tiny[0.520,\,0.712]}}}$ & $\mathbf{9.6}$ ($41.4\%$) \\
\bottomrule
\end{tabular}
\end{threeparttable}
\end{table*}

\begin{figure*}
    \centering
    \includegraphics[width=0.65\linewidth]{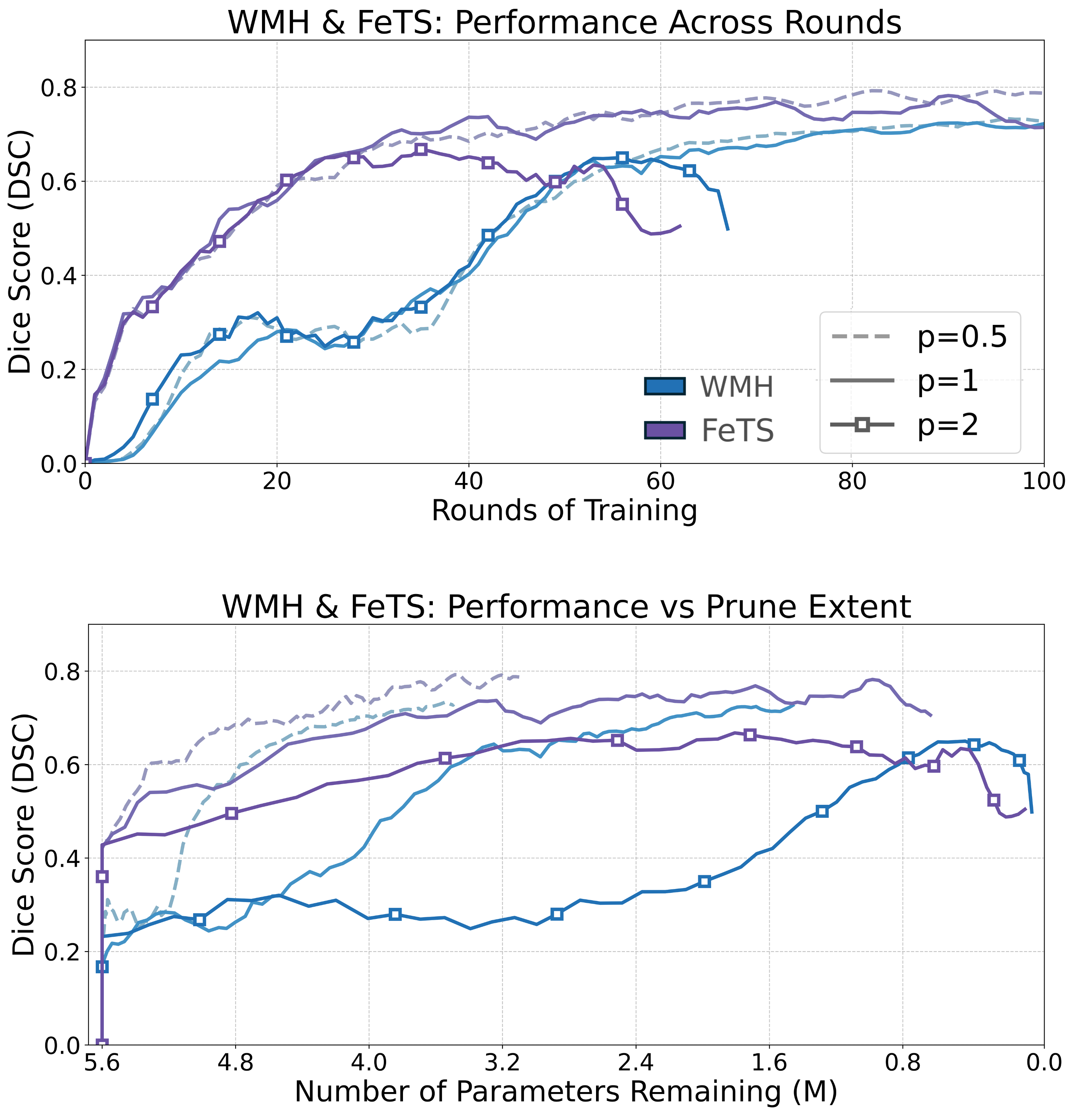}
    \caption{Effect of the varying pruning percentage \textit{p} on segmentation performance across rounds (top panel) and parameter efficiency (bottom panel) during FedASAP training, for WMH and FeTS datasets. 
    The sharp initial rise in Dice scores (right subplots) corresponds to the warm-up period of training, during which no pruning is applied and the model rapidly improves. The steep declines at the end indicate excessive pruning, leaving insufficient capacity for accurate segmentation.}
    \label{fig:effect_fig}
\end{figure*}

\subsection{Effect of the pruning parameter p} As shown in Figure~\ref{fig:effect_fig}, increasing \textit{p} (e.g. \textit{p}=2) results in more aggressive pruning and faster reduction in model size, but this comes at the cost of reduced segmentation accuracy. On the other hand, lower values of \textit{p} (e.g. \textit{p}=0.5) are more conservative, preserving performance but offering limited compression. Interestingly, \textit{p}=1 emerges as a balanced choice—achieving comparable segmentation performance to \textit{p}=0.5 while retaining fewer parameters throughout training. This suggests that intermediate values of \textit{p} offer a favorable trade-off between compactness and accuracy for both WMH and FeTS clients. These results underscore the importance of selecting an optimal \textit{p} to balance convergence speed, model compression, and segmentation performance.

\section{Discussion} 
In this work, we present FedASAP, a novel federated learning framework for brain MRI lesion segmentation that incorporates activation statistics-driven structured adaptive filter pruning to address the dual challenges of data heterogeneity and computational efficiency in federated settings. By leveraging a two-stage adaptive pruning approach, comprising initial score-based screening followed by GLM-based filter selection, FedASAP adjusts pruning strategies across clients based on lesion characteristics, optimizing segmentation performance for both small lesions (WMH) and large lesions (FeTS). By performing pruning entirely client-side, FedASAP enables personalized model creation tailored to each client's unique data distribution, making it robust for real-world non-IID scenarios.
FedASAP consistently outperforms unstructured federated-pruning baselines (FedDST, AutoFLIP, PruneFL) (Table~\ref{tab:wmh_fets_metrics}), highlighting the effectiveness of two-stage fine-grained pruning in balancing performance and computational efficiency. Unlike other methods that require a predefined sparsity level, FedASAP uses incremental pruning with no \emph{a priori} sparsity setting, allowing weights to adapt as sparsity increases. This helps maintain stable gradients and prevents sudden loss spikes that could harm generalization. FedASAP also incorporates activation-based statistics into the pruning decision, prioritizing important filters in an explainable manner. This approach was particularly useful in observing that, in deeper architectures, excessive representational capacity may not be beneficial for small-lesion tasks. For instance, for WMH, GBFS refines the selection to recover or improve accuracy while achieving higher sparsity, a pattern that holds across different depths.
Beyond accuracy and sparsity, FedASAP delivers tangible run-time and memory benefits. This confirms that unstructured sparsity alone does not translate into practical computational or memory reductions on standard hardware/software stacks, as illustrated in  Figure~\ref{fig:metrics-comparison}. While AutoFLIP demands extensive exploration training and FedDST requires additional local retraining epochs (notably in our WMH experiments), our incremental pruning strategy explicitly and progressively reduces model size during training, leading to a gradual decrease in computational overhead. This behavior contrasts with unstructured, sparsity-inducing methods, which often maintain a nearly constant execution cost on standard hardware despite their theoretical sparsity.

Additionally, FedASAP incurs only modest extra communication overhead relative to PruneFL, FedDST, and AutoFLIP: its upload cost scales with the total number of filters rather than with the number of clients $n$ or rounds $T$ as reported in Table~\ref{tab:comm_compare}. However, because pruning is progressive, the model remains large in early rounds, so communication costs are front-loaded until the model footprint is substantially reduced through pruning.
The proposed approach can be integrated directly with standard FL frameworks (FedAvg, FedProx, MOON, etc.) without modifying core aggregation logic, enabling seamless plug‐and‐play deployment. 

The addition of LAT modules emphasizes that the pruning strategy could effectively reflect lesion characteristics. For WMH segmentation dominated by small lesions, while PLAT and NLAT provide similar performances, NLAT results in a much larger network than PLAT (2.9M vs. 1.3M parameters). This shows that, even though NLAT is designed primarily to retain filters in deeper layers, due to small lesion size, retention of fine‐detail filters in shallow layers is also inevitable for small‐lesion detection. In contrast, in large-lesion dataset like FeTS, NLAT outperforms PLAT, demonstrating the need to preserve coarse, structural features in central layers. FedASAP provides a better, more explainable activation statistics‐driven pruning. This avoids the need for manually choosing the filter proportion in shallow and deep layers as done in LAT, which would be particularly useful in the cases where the lesion characteristics of a client are not known in prior. Note that in addition to features specified in Section~\ref{sssec:features}, a few features including sparsity, median absolute deviation were tried but later removed due to their negligible contribution. Hence, for other segmentation tasks, based on lesion characteristics, additional relevant features can be added to train GLM, making FedASAP an adaptive and flexible framework.

Overall, FedASAP improves accuracy, reduces runtime and memory costs in the presence of lesion heterogeneity. However, a limitation of FedASAP is that the performance degrades under the pooled setup (mixture of WMH and FeTS within a client). This is likely due to the larger domain shift between clients and mixed lesion attributes that could affect the decision of GLM-classifier to retain finer vs. coarser filters in deeper and shallow layers. Future directions include incorporating additional radiomics-based activation statistics into the GLM framework to better distinguish informative filters and extending the approach to multi-modal scenarios (e.g., FLAIR + T1) by leveraging cross-modal activations to guide filter removal to improve FedASAP's adaptability and performance.

\section{Conclusions}
We propose FedASAP, a two-stage, fine-grained structural adaptive pruning framework for personalized federated medical image segmentation that optimizes segmentation accuracy and model compactness. Our method shows superior segmentation performance compared to existing federated learning approaches while reducing model size, achieving parameter reductions of up to $73.0\%$. Compared to state-of-the-art unstructured pruning approaches, FedASAP requires minimal communication overhead while delivering gains in training time, memory usage, and computational load unattainable through unstructured approaches. FedASAP is suitable for deployment in resource-constrained medical institutions, and can be integrated with standard FL frameworks. Code will be made available at \url{https://github.com/Tarun2201/FedASAP_development}.

\printcredits

\section*{Declaration of competing interest}
The authors declare that they have no known competing financial interests or personal relationships that could have appeared to influence the work reported in this paper. The authors have patent titled `A method and system for two staged adaptive structured pruning in federated medical image segmentation' pending to Indian Institute of Science, Bangalore, India.

\section*{Ethics statement}
For the Federated Tumour Segmentation (FeTS) Dataset, the dataset includes multi-institutional data. Informed consent was obtained from all subjects at their respective institutions, and the protocol for releasing the data was approved by the institutional review board of the data contributing institution.
For the WMH dataset, the multi-institutional data is available as part of MICCAI 2017 WMH segmentation Challenge and the study was approved by the institutional review boards of the Amsterdam UMC and the UMCU (approval number 14-083/C). Since both datasets are publicly available, for this study, no additional ethics approval is required.

\section*{Declaration of Generative AI and AI-assisted technologies in the
writing process}
This manuscript was written entirely by the authors without the use
of any large language models (LLMs) or artificial intelligence tools. All
content, including analysis, interpretations, and conclusions, is solely
the product of the authors’ research and intellectual effort.

\section*{Acknowledgments}
This work was supported by DBT/Wellcome Trust India Alliance Fellowship [IA/E/22/1/506763]. This work was also supported in part by a grant from the Council of Scientific \& Industrial Research (CSIR) under its ASPIRE (Women Scientist Scheme) program [25WS(013)/2023-24/EMR-II/ASPIRE], in part by Start-up Research Grant [SRG/2023/001406] from the Science and Engineering Research Board, India and in part by Siemens Healthineers-CDS Collaborative Laboratory of Artificial Intelligence in Precision Medicine, India. VS is also supported by Pratiksha Trust, Bangalore, India [FG/PTCH-23-1004] and the Seed Research Grant [IE/RERE-22-0583] from the Indian Institute of Science, India.
\bibliographystyle{unsrtnat}

\bibliography{ref}

\end{document}